\documentclass[sigconf]{acmart}
\AtBeginDocument{
  }

\usepackage{enumitem}
\usepackage{amsmath}

\usepackage{graphicx}
\usepackage{booktabs}
\usepackage{microtype}
\usepackage{url}
\usepackage{multirow}
\usepackage{stfloats}

\usepackage{tikz, subcaption}
\usetikzlibrary{arrows.meta, positioning}

\usepackage{listings}
\usepackage[T1]{fontenc}
\copyrightyear{2026}
\acmYear{2026}
\setcopyright{cc}
\setcctype{by}
\acmConference[WWW '26]{Proceedings of the ACM Web Conference 2026}{April 13--17, 2026}{Dubai, United Arab Emirates}
\acmBooktitle{Proceedings of the ACM Web Conference 2026 (WWW '26), April 13--17, 2026, Dubai, United Arab Emirates}
\acmPrice{}
\acmDOI{10.1145/3774904.3792955}
\acmISBN{979-8-4007-2307-0/2026/04}

\newcommand{\mysubhead}[1]{\noindent\textbf{#1}}

\begin{document}

\title{Retrieval Collapses When AI Pollutes the Web}

\author{Hongyeon Yu}
\affiliation{
  \institution{NAVER Corp.}
  \city{Bellevue}
  \state{WA}
  \country{USA}
}
\email{hongyeon.yu@navercorp.com}

\author{Dongchan Kim}
\affiliation{
  \institution{NAVER Corp.}
  \city{Bellevue}
  \state{WA}
  \country{USA}
}
\email{dongchan.usa@gmail.com}

\author{Young-Bum Kim}
\affiliation{
  \institution{NAVER Corp.}
  \city{Bellevue}
  \state{WA}
  \country{USA}
}
\email{youngbum.kim@navercorp.com}

\renewcommand{\shortauthors}{Hongyeon Yu, Dongchan Kim, and Young-Bum Kim}
\begin{abstract}
The rapid proliferation of AI-generated content on the Web presents a structural risk to information retrieval, as search engines and Retrieval-Augmented Generation (RAG) systems increasingly consume evidence produced by the Large Language Models (LLMs). We characterize this ecosystem-level failure mode as Retrieval Collapse, a two-stage process where (1) AI-generated content dominates search results, eroding source diversity, and (2) low-quality or adversarial content infiltrates the retrieval pipeline. We analyzed this dynamic through controlled experiments involving both high-quality SEO-style content and adversarially crafted content. In the SEO scenario, a 67\% pool contamination led to over 80\% exposure contamination, creating a homogenized yet deceptively healthy state where answer accuracy remains stable despite the reliance on synthetic sources. Conversely, under adversarial contamination, baselines like BM25 exposed $\sim$19\% of harmful content, whereas LLM-based rankers demonstrated stronger suppression capabilities. These findings highlight the risk of retrieval pipelines quietly shifting toward synthetic evidence and the need for retrieval-aware strategies to prevent a self-reinforcing cycle of quality decline in Web-grounded systems.
\end{abstract}



\maketitle

\section{Introduction}

The rapid proliferation of Large Language Models (LLMs) has fundamentally transformed the landscape of Web content creation~\cite{spennemann2025syntheticcontent}. Although this shift offers scalability in information production, it introduces a critical structural vulnerability for search engines and Retrieval-Augmented Generation (RAG) systems~\cite{lewis2020retrieval, izacard2021fid}. These systems increasingly consume evidence that is itself generated by the very models they rely on, creating a self-referential cycle. While similar phenomena have been studied in model training as \textit{model collapse}~\cite{shumailov2024modelcollapse, alemohammad2024selfconsuming}, the implications for the \textit{retrieval} ecosystem remain underexplored.

We characterize this ecosystem-level failure mode as Retrieval Collapse, a two-stage degradation process. The first stage, \textit{Dominance and Homogenization}, occurs when high-quality, SEO-optimized synthetic content captures the top search results, drastically reducing source diversity. This stage is particularly insidious because it creates a \textit{deceptively healthy} state: surface-level answer quality may remain stable due to the fluency of LLM outputs, masking the underlying erosion of information provenance. The second stage, \textit{Pollution and System Corruption}, emerges when low-quality or adversarial content infiltrates the pipeline. In this phase, malicious actors can exploit ranking algorithms to inject misleading information, undermining the factual integrity of downstream RAG systems.

To investigate these risks, we constructed a controlled experimental environment using the MS MARCO dataset, simulating two distinct contamination scenarios: (1) a high-quality SEO pool and (2) an adversarial abuse pool. Our findings reveal a contrast in system behavior. Under the SEO scenario, a Pool Contamination Rate of 67\% led to an Exposure Contamination Rate exceeding 80\% across rankers, confirming a rapid shift toward homogenized synthetic evidence~\cite{dai2024sourcebias}. Conversely, under the adversarial scenario, while LLM-based rankers showed resilience, widely deployed baselines like BM25 exhibited significant vulnerability, exposing approximately 19\% of harmful content at the same contamination level.

This study highlights that current retrieval pipelines can quietly shift toward synthetic or harmful evidence without immediate signs of quality collapse. Our contributions are threefold:
\begin{itemize}
    \item We formally conceptualize Retrieval Collapse as a structural failure mode distinct from training-time model collapse.
    \item We provide empirical evidence of contamination dynamics, quantifying how 67\% pool contamination translates to over 80\% exposure in SEO settings and $\sim$19\% vulnerability in BM25 under adversarial attacks.
    \item We underscore the necessity for retrieval-aware ranking strategies that go beyond topical relevance to preserve ecosystem diversity and trust.
\end{itemize}

\section{Background and Related Work}

While generative models degrade when trained on their own outputs, a phenomenon termed \textit{model collapse}~\cite{shumailov2024modelcollapse, alemohammad2024selfconsuming}, prior research typically focuses on closed training loops.
In contrast, we address structural risks during retrieval, where heterogeneous generators and ranking systems shape the exposure landscape.
Recent studies have begun to explore similar feedback loops in Open Domain Question Answering (ODQA)~\cite{chen-etal-2024-spiral} and RAG corpus poisoning~\cite{su2025corpus, ying2024safebench}, where adversarial actors inject malicious documents.
However, our work differentiates itself by analyzing the \textit{ecosystem-level} shift caused by mass-produced SEO content rather than isolated adversarial attacks.

With the proliferation of AI-generated text~\cite{spennemann2025syntheticcontent}, challenges in attribution~\cite{li2024mage, sadasivan2023impossible} and pre-training data quality~\cite{dodge2021documenting, birhane2021multimodal} have intensified.
Unlike traditional keyword spam~\cite{gyongyi2005webspam}, modern synthetic content is semantically coherent, allowing it to blend into ranking systems and propagate through pipelines as authoritative evidence~\cite{dai2024bias, dai2024sourcebias, zhou2025sourceecho, xu2024invisiblerelevance}.
We extend these findings by quantifying distinct contamination dynamics driven by SEO-style versus adversarial content.
Although provenance techniques like datasheets~\cite{gebru2021datasheets} and watermarking~\cite{kirchenbauer2023watermark} offer document-level mechanisms, they do not fully address the ecosystem degradation where the collective presence of synthetic documents reshapes ranking behavior.

\section{Experimental Methodology}
\label{sec:methodology}

We construct a controlled environment to evaluate how different forms of AI-generated content propagate through retrieval pipelines and induce the two-stage dynamics of Retrieval Collapse.\footnote{The source code used in our experiments is publicly available at \url{https://github.com/dongchankim-io/retrieval-collapse} to support reproducibility.}

\subsection{Dataset}

Our experiments utilize 1,000 query-answer pairs randomly sampled from the \textsc{MS MARCO} passage ranking benchmark.
MS MARCO provides diverse open-domain queries paired with human-validated reference answers, which we use both to ground retrieval tasks and to evaluate factual correctness.
We next construct three document pools used as reference sources for answer generation.

\mysubhead{Original Pool.}
For each MS MARCO query, we retrieve ten Web documents using the Google Search API.
As these are top-ranked results from a major search engine, they inherently reflect current industry-standard SEO optimization.
We follow each URL to extract full article content, producing an Original Pool of $N = 10{,}000$ documents.
To determine factual validity, each document is verified by an LLM Judge against the ground truth, yielding a Micro Correct Rate of 51.69\%.

\mysubhead{SEO Pool (High-Quality Synthetic Content).}
To simulate content farms, we generate 20 synthetic documents per query using GPT-5-nano.
We generate a larger pool than the original set to ensure sufficient unique synthetic content is available for the 20-round cumulative contamination simulation.
Each SEO document is created by sampling a \emph{random combination} of the Original Pool documents and synthesizing them into a coherent article.
This models the realistic scenario where non-expert users utilize LLMs to aggregate and refine search results into optimized pages.
The SEO Pool exhibits a Micro Correct Rate of 66.79\%, reflecting the "deceptively healthy" Stage~1 \textit{Dominance}.

\mysubhead{Abuse Pool (Adversarial Synthetic Content).}
To simulate malicious contamination, the Abuse Pool is constructed \emph{without} using Original documents.
Each document is created via a two-step adversarial pipeline: (1) a \emph{Clickbait/SEO Generator} produces an engaging draft, and (2) a \emph{Misleading Content Rewriter} adversarially replaces factual entities with plausible but incorrect alternatives.
This simulates Stage~2 pollution. The Abuse Pool has a Micro Correct Rate of 38.44\%.

\subsection{LLM Modules and Simulation}

Our retrieval environment consists of four components instantiated using the GPT-5 family.
We employ different model sizes to balance realistic simulation constraints against evaluation rigor.
The \textbf{Content Generator} (GPT-5-nano) produces synthetic documents. We selected the `nano` variant to simulate the economic reality of content farms, where low-latency, low-cost models are preferred for mass production.
The \textbf{LLM Ranker} (GPT-5-nano) performs semantic re-ranking, and the \textbf{LLM Answerer} (GPT-5-nano) synthesizes final responses.
Crucially, the \textbf{LLM Judge} (GPT-5-mini) employs a more capable model than the generators. This ensures a high-quality upper bound for correctness evaluation and reduces the risk of self-confirmation bias where a model might preferentially rate its own output highly.

To simulate the gradual growth of AI-generated content, we run a 20-round contamination process.
In each round, one synthetic document is cumulatively added to the fixed set of ten Original documents per query, increasing the AI ratio from 0\% (Round~0) to 66.7\% (Round~20).

\mysubhead{Prompting Strategy.}
To ensure the SEO Pool mimicked realistic content farms, we utilized a role-playing prompt instructing the generator to ``act as an SEO specialist,'' explicitly integrating high-IDF keywords extracted from the Original Pool to maximize retrieval likelihood.
Conversely, for the Abuse Pool, the prompt emphasized ``preserving surface-level fluency while altering specific named entities and numerical facts,'' thereby creating documents that pass statistical filters while degrading information integrity.

\subsection{Evaluation Metrics}

We evaluate how contamination propagates through the retrieval stack using three metrics: \textbf{Pool Contamination Rate (PCR)}, the fraction of AI-generated documents in the full pool; \textbf{Exposure Contamination Rate (ECR)}, the fraction appearing in the top-10 retrieved results; and \textbf{Citation Contamination Rate (CCR)}, the fraction explicitly cited by the LLM Answerer. ECR reflects contamination entering the RAG pipeline, whereas CCR captures the evidence actually used to generate an answer.

To assess user-facing impact, we use two standard retrieval metrics: \textbf{Precision@10 (P@10)} measures the proportion of retrieved documents that are factually correct (i.e., labeled as correct based on the ground truth). \textbf{Answer Accuracy (AA)} evaluates the validity of the final response. Specifically, we employ the LLM Judge to verify whether the answer generated by the LLM Answerer is semantically consistent with the MS MARCO ground-truth answer as reference.

\begin{table*}[!t]
  \centering
  \caption{Contamination statistics across Scenarios 1 and 2 under both BM25 and LLM Rankers.}
  \label{tab:unified_all}
  \begin{subtable}[b]{0.48\textwidth}
    \caption{Scenario 1}
    \centering
    \begin{tabular}{c|c|c|c|c|c}
      \toprule
      \textbf{Ranker} & \textbf{Round} & \textbf{PCR} & \textbf{ECR} & \textbf{CCR} & \textbf{AA} \\
      \midrule
      \multirow{4}{*}{BM25}
      & 0  & 0.00 & 0.0000 & 0.0000 & 0.6817 \\
      & 5  & 0.33 & 0.4291 & 0.5613 & 0.6947 \\
      & 10 & 0.50 & 0.6809 & 0.7683 & 0.6911 \\
      & 20 & 0.67 & 0.8095 & 0.8695 & 0.6768 \\
      \midrule
      \multirow{4}{*}{LLM}
      & 0  & 0.00 & 0.0000 & 0.0000 & 0.6841 \\
      & 5  & 0.33 & 0.4508 & 0.5265 & 0.7019 \\
      & 10 & 0.50 & 0.7602 & 0.7853 & 0.7089 \\
      & 20 & 0.67 & 0.7998 & 0.8170 & 0.7023 \\
      \bottomrule
    \end{tabular}
    \label{tab:seo_stats}
  \end{subtable}
  \hfill
  \begin{subtable}[b]{0.48\textwidth}
    \caption{Scenario 2}
    \centering
    \begin{tabular}{c|c|c|c|c|c}
      \toprule
      \textbf{Ranker} & \textbf{Round} & \textbf{PCR} & \textbf{ECR} & \textbf{CCR} & \textbf{AA} \\
      \midrule
      \multirow{4}{*}{BM25}
      & 0  & 0.00 & 0.0000 & 0.0000 & 0.6817 \\
      & 5  & 0.33 & 0.1395 & 0.0000 & 0.6792 \\
      & 10 & 0.50 & 0.2434 & 0.0010 & 0.6678 \\
      & 20 & 0.67 & 0.1897 & 0.0050 & 0.6561 \\
      \midrule
      \multirow{4}{*}{LLM}
      & 0  & 0.00 & 0.0000 & 0.0000 & 0.6841 \\
      & 5  & 0.33 & 0.0002 & 0.0000 & 0.6663 \\
      & 10 & 0.50 & 0.0001 & 0.0000 & 0.6904 \\
      & 20 & 0.67 & 0.0009 & 0.0000 & 0.6794 \\
      \bottomrule
    \end{tabular}
    \label{tab:abuse_stats}
  \end{subtable}
\end{table*}

\section{Results and Analysis}

\subsection{Baseline Ranker}
We first evaluate ranker performance on the pristine \textit{Original Pool}. The LLM Ranker achieves strong retrieval quality ($nDCG@5=0.6251$), slightly outperforming the scalable BM25 Ranker baseline ($nDCG@5=0.6125$).

\subsection{Scenario~1: Dominance and Homogenization}

We evaluate this effect in Scenario 1, which examines how SEO-style synthetic documents reshape retrieval (Figure~\ref{fig:seo_bm25}; Table~\ref{tab:seo_stats}).

\mysubhead{Contamination Convergence.}
Across both rankers, exposure contamination is significantly amplified relative to pool contamination. Under BM25, ECR surpasses 68\% by Round~10 ($PCR = 50\%$) and exceeds 80\% by Round~20 ($PCR= 67\%$). The LLM Ranker exhibits a stronger preference, with ECR reaching 76\% at Round~10 and remaining consistently higher than that of BM25. This pattern shows that SEO-optimized content disproportionately activates ranking signals, causing both models to converge rapidly toward synthetic-dominated evidence.

\mysubhead{Factual Stability vs.\ Diversity Collapse.}
Despite this dramatic shift in retrieved evidence, AA remains stable or slightly improves (from roughly 68\% to 70\%). Because SEO documents are high-quality and topically aligned, retrieval appears healthy when measured solely by accuracy. However, nearly all retrieved evidence is synthetic, indicating a severe collapse in source diversity. This divergence, characterized by stable accuracy despite collapsing diversity, reveals a structurally brittle retrieval pipeline: the system performs well in aggregate metrics while quietly losing its grounding in human-written content.

Overall, high-quality synthetic content not only integrates seamlessly into retrieval pipelines but actively overwhelms ranking signals, leading both BM25 and LLM Rankers to rely almost exclusively on AI-generated evidence.

\subsection{Scenario~2: Pollution and System Corruption}

\begin{figure*}[t]
  \centering
  \begin{subfigure}[b]{0.48\textwidth}
    \includegraphics[width=\linewidth]{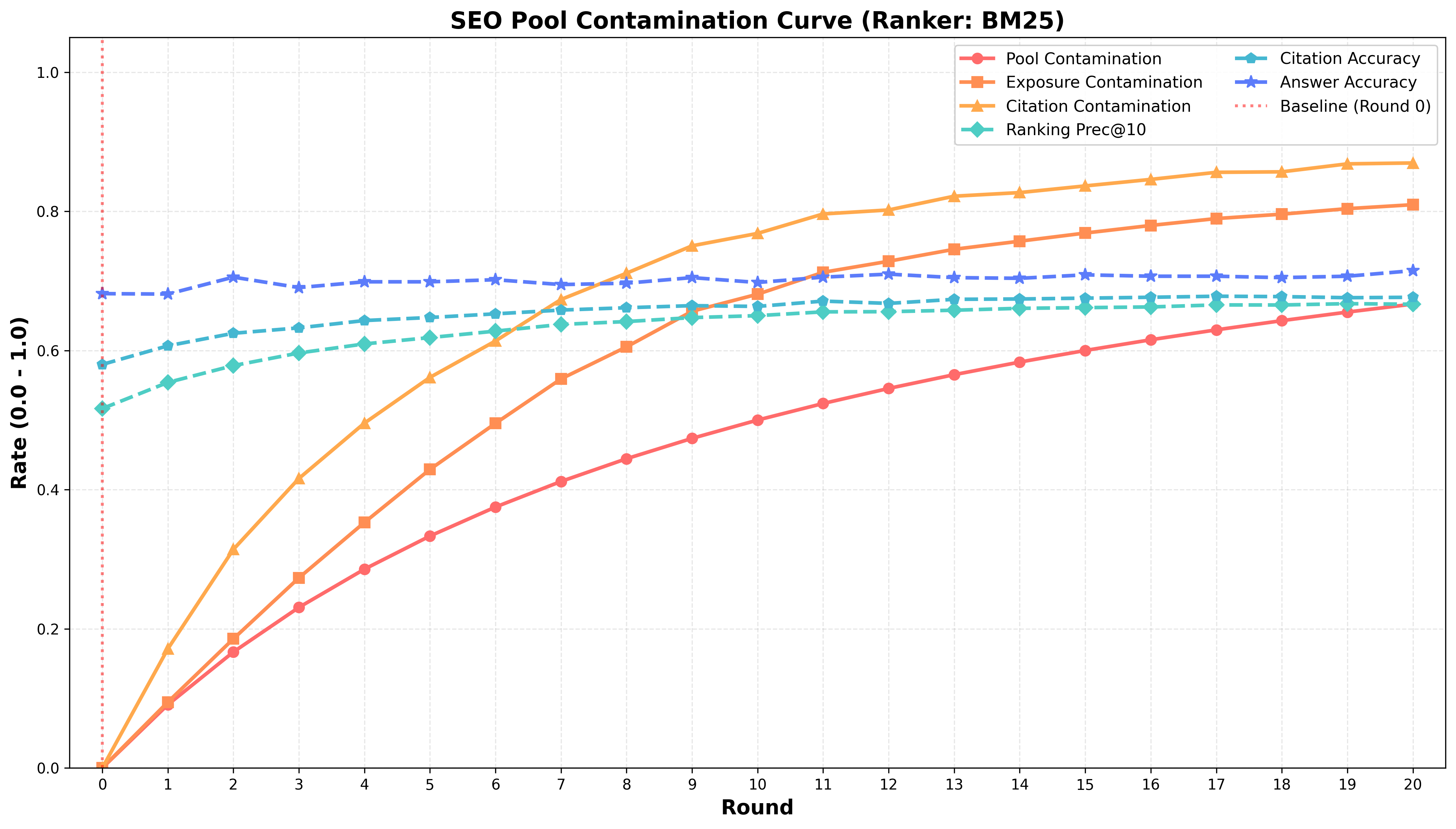}
    \caption{Scenario 1}
    \label{fig:seo_bm25}
  \end{subfigure}
  \hfill
  \begin{subfigure}[b]{0.48\textwidth}
    \includegraphics[width=\linewidth]{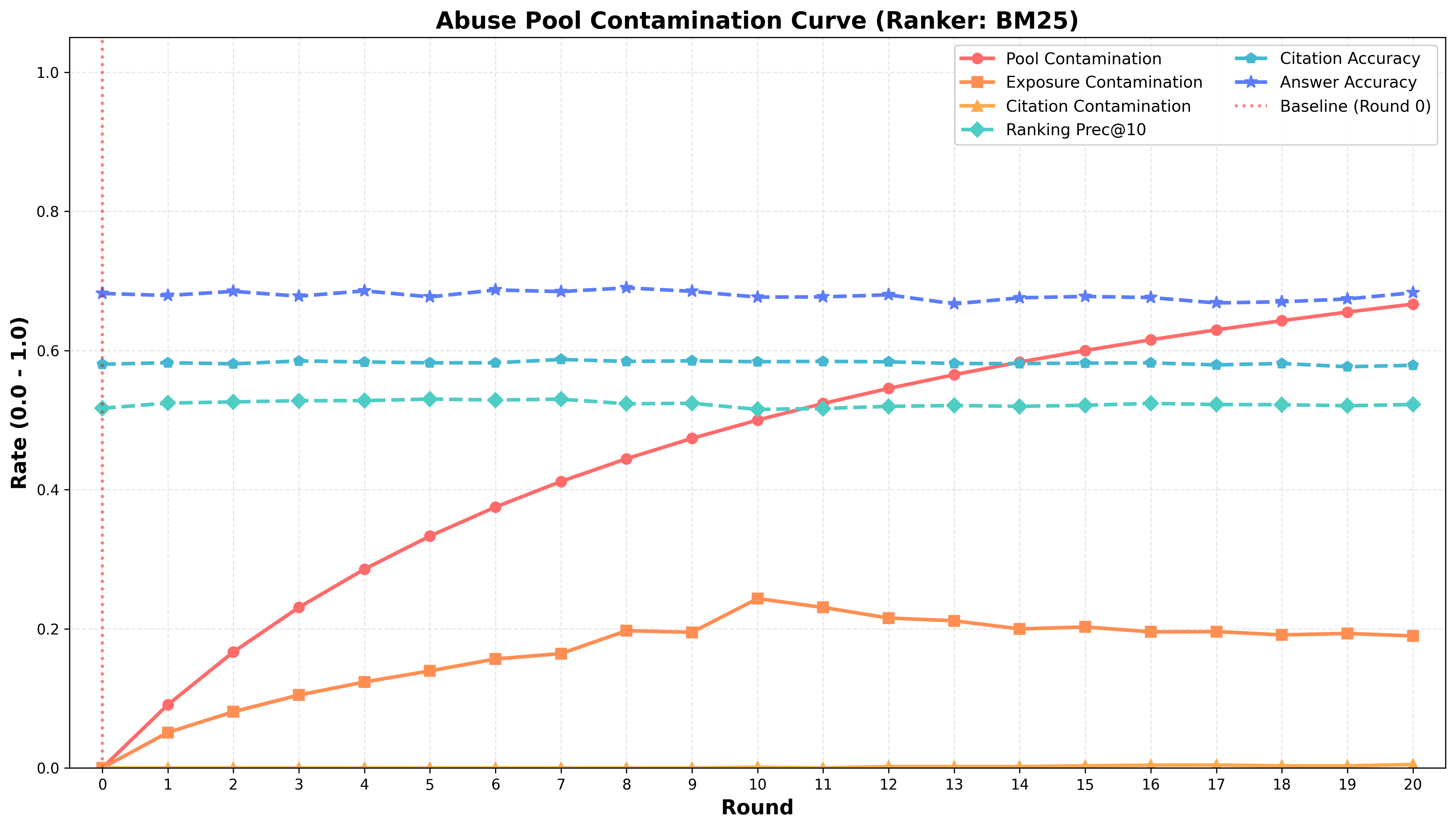}
    \caption{Scenario 2}
    \label{fig:abuse_bm25}
  \end{subfigure}
  \caption{Contamination dynamics across SEO and adversarial settings under BM25.}
  \label{fig:main_results}
\end{figure*}

Scenario~2 investigates the second stage of Retrieval Collapse using the adversarial Abuse Pool. This scenario highlights a critical divergence in ranker behavior compared to Scenario~1 (Figure~\ref{fig:abuse_bm25}; Table~\ref{tab:abuse_stats}).

\mysubhead{Robustness of LLM Rankers vs.\ Vulnerability of BM25.}
A striking finding is the LLM Ranker's resilience. Unlike in Scenario~1, where it was overwhelmed by SEO content, LLM Ranker successfully detects and suppresses lower-quality adversarial documents, maintaining ECR near zero. However, relying exclusively on computationally expensive LLM Rankers is often impractical for large-scale systems. The scalable BM25 Ranker, while performing better than in the SEO setting, still exhibits a critical vulnerability: it allows approximately 19--24\% of abusing documents to infiltrate the top-10 results. This represents a significant structural risk for standard retrieval pipelines.

\mysubhead{Apparent Stability vs.\ Underlying Degradation.}
Despite the retrieval breach under BM25, AA appears relatively stable. This is primarily because the LLM Answerer successfully suppressed citation corruption, acting as a final filter against adversarial content. However, this stability masks a critical vulnerability at the retrieval layer. Furthermore, a comparative analysis reveals that answer accuracy in Scenario~2 consistently underperforms compared to Scenario~1. While Scenario~1 saw AA maintained or even improved (reaching up to 70\% with LLM Rankers) due to the high quality of SEO content, Scenario~2 exhibits a decline in answer quality relative to the SEO setting. Specifically for BM25, AA drops below the original baseline ($68\% \to 66\%$). This confirms that regardless of the ranker, adversarial pollution in the retrieval stage negatively impacts end-to-end performance, with the degradation being most severe when relying on lightweight retrievers.

\section{Implications and Conclusion}

We formally introduced and empirically validated Retrieval Collapse, a two-stage structural issue where synthetic content first achieves \textit{Dominance} and subsequently facilitates \textit{System Corruption}.
Our Scenario~1 findings expose a critical loss in source diversity, introducing extreme brittleness where high accuracy masks ecosystem decay.
Scenario~2 demonstrates that scalable baselines like BM25 are critically vulnerable to adversarial pollution ($19\%$ exposure), whereas LLM-based rankers offer resilience but at high computational cost.

By establishing the framework of Retrieval Collapse, this work lays the foundation for understanding how synthetic content reshapes information retrieval.
To mitigate these risks, we propose a shift toward Defensive Ranking strategies that jointly optimize relevance, factuality, and provenance~\cite{joachims2017ranking}.

Effective mitigation requires moving beyond retrieval-time re-ranking to implement scalable, \textit{ingestion-stage} safeguards, such as deploying lightweight ``perplexity filters'' or ``provenance graphs'' that flag content with high fluency but low attribution density before it enters the index.
Furthermore, as Agentic AI begins to autonomously publish content, defense mechanisms must evolve from static text analysis to behavioral fingerprinting, identifying and isolating agents that systematically produce high-entropy, low-factuality streams.

\mysubhead{Limitations.} 
We acknowledge that our metrics (PCR, ECR, CCR) are descriptive decompositions intended to characterize dynamics rather than novel theoretical measures. Additionally, our SEO simulation assumes generic LLM-based optimization rather than expert-level adversarial engineering.
Future work should explore agentic threats where autonomous AI explicitly attempts to manipulate retrieval rankings, and validate these findings in live, large-scale web environments.


\bibliographystyle{ACM-Reference-Format}
\bibliography{references}

@article{shumailov2024modelcollapse,
	author = {Shumailov, Ilia and Shumaylov, Zakhar and Zhao, Yiren and Papernot, Nicolas and Anderson, Ross and Gal, Yarin},
	date = {2024/07/01},
	doi = {10.1038/s41586-024-07566-y},
	id = {Shumailov2024},
	isbn = {1476-4687},
	journal = {Nature},
	number = {8022},
	pages = {755--759},
	title = {AI models collapse when trained on recursively generated data},
	url = {https://doi.org/10.1038/s41586-024-07566-y},
	volume = {631},
	year = {2024}
}

@misc{alemohammad2024selfconsuming,
  title = {Self-Consuming Generative Models Go MAD}, 
  author = {Sina Alemohammad and Josue Casco-Rodriguez and Lorenzo Luzi and Ahmed Imtiaz Humayun and Hossein Babaei and Daniel LeJeune and Ali Siahkoohi and Richard G. Baraniuk},
  year = {2023},
  eprint = {2307.01850},
  archivePrefix = {arXiv},
  primaryClass = {cs.LG},
  url = {https://arxiv.org/abs/2307.01850}, 
}

@misc{spennemann2025syntheticcontent,
  title = {Delving into: the quantification of AI-generated content on the internet (synthetic data)}, 
  author = {Dirk HR Spennemann},
  year = {2025},
  eprint = {2504.08755},
  archivePrefix = {arXiv},
  primaryClass = {cs.IR},
  url = {https://arxiv.org/abs/2504.08755}, 
}

@article{ying2024safebench,
  author = {Ying, Zonghao and Liu, Aishan and Liang, Siyuan and Huang, Lei and Guo, Jinyang and Zhou, Wenbo and Liu, Xianglong and Tao, Dacheng},
  title = {SafeBench: A Safety Evaluation Framework for Multimodal Large Language Models},
  year = {2025},
  issue_date = {Jan 2026},
  publisher = {Kluwer Academic Publishers},
  address = {USA},
  volume = {134},
  number = {1},
  issn = {0920-5691},
  url = {https://doi.org/10.1007/s11263-025-02613-1},
  doi = {10.1007/s11263-025-02613-1},
  journal = {Int. J. Comput. Vision},
  month = dec,
  numpages = {25}
}

@inproceedings{dodge2021documenting,
  title = "Documenting Large Webtext Corpora: A Case Study on the Colossal Clean Crawled Corpus",
  author = "Dodge, Jesse and Sap, Maarten and Marasovi{\'c}, Ana and Agnew, William and Ilharco, Gabriel and Groeneveld, Dirk and Mitchell, Margaret and Gardner, Matt",
  editor = "Moens, Marie-Francine and Huang, Xuanjing and Specia, Lucia and Yih, Scott Wen-tau",
  booktitle = "Proceedings of the 2021 Conference on Empirical Methods in Natural Language Processing",
  month = nov,
  year = "2021",
  address = "Online and Punta Cana, Dominican Republic",
  publisher = "Association for Computational Linguistics",
  url = "https://aclanthology.org/2021.emnlp-main.98/",
  doi = "10.18653/v1/2021.emnlp-main.98",
  pages = "1286--1305",
}

@misc{birhane2021multimodal,
  title = {Multimodal datasets: misogyny, pornography, and malignant stereotypes}, 
  author = {Abeba Birhane and Vinay Uday Prabhu and Emmanuel Kahembwe},
  year = {2021},
  eprint = {2110.01963},
  archivePrefix = {arXiv},
  primaryClass = {cs.CY},
  url = {https://arxiv.org/abs/2110.01963}, 
}

@inproceedings{li2024mage,
  title = "{MAGE}: Machine-generated Text Detection in the Wild",
  author = "Li, Yafu and Li, Qintong and Cui, Leyang and Bi, Wei and Wang, Zhilin and Wang, Longyue and Yang, Linyi and Shi, Shuming and Zhang, Yue",
  editor = "Ku, Lun-Wei and Martins, Andre and Srikumar, Vivek",
  booktitle = "Proceedings of the 62nd Annual Meeting of the Association for Computational Linguistics (Volume 1: Long Papers)",
  month = aug,
  year = "2024",
  address = "Bangkok, Thailand",
  publisher = "Association for Computational Linguistics",
  url = "https://aclanthology.org/2024.acl-long.3/",
  doi = "10.18653/v1/2024.acl-long.3",
  pages = "36--53",
}

@inproceedings{dai2024sourcebias,
  series = {KDD '24},
  title = {Neural Retrievers are Biased Towards LLM-Generated Content},
  url = {http://dx.doi.org/10.1145/3637528.3671882},
  DOI = {10.1145/3637528.3671882},
  booktitle = {Proceedings of the 30th ACM SIGKDD Conference on Knowledge Discovery and Data Mining},
  publisher = {ACM},
  author = {Dai, Sunhao and Zhou, Yuqi and Pang, Liang and Liu, Weihao and Hu, Xiaolin and Liu, Yong and Zhang, Xiao and Wang, Gang and Xu, Jun},
  year = {2024},
  month = aug,
  pages = {526–537},
  collection = {KDD '24}
}

@inproceedings{lewis2020retrieval,
  author = {Lewis, Patrick and Perez, Ethan and Piktus, Aleksandra and Petroni, Fabio and Karpukhin, Vladimir and Goyal, Naman and K\"{u}ttler, Heinrich and Lewis, Mike and Yih, Wen-tau and Rockt\"{a}schel, Tim and Riedel, Sebastian and Kiela, Douwe},
  title = {Retrieval-augmented generation for knowledge-intensive NLP tasks},
  year = {2020},
  isbn = {9781713829546},
  publisher = {Curran Associates Inc.},
  address = {Red Hook, NY, USA},
  booktitle = {Proceedings of the 34th International Conference on Neural Information Processing Systems},
  articleno = {793},
  numpages = {16},
  location = {Vancouver, BC, Canada},
  series = {NIPS '20}
}

@inproceedings{izacard2021fid,
  title = "Leveraging Passage Retrieval with Generative Models for Open Domain Question Answering",
  author = "Izacard, Gautier and Grave, Edouard",
  editor = "Merlo, Paola and Tiedemann, Jorg and Tsarfaty, Reut",
  booktitle = "Proceedings of the 16th Conference of the European Chapter of the Association for Computational Linguistics: Main Volume",
  month = apr,
  year = "2021",
  address = "Online",
  publisher = "Association for Computational Linguistics",
  url = "https://aclanthology.org/2021.eacl-main.74/",
  doi = "10.18653/v1/2021.eacl-main.74",
  pages = "874--880",
}

@inproceedings{joachims2017ranking,
  author = {Joachims, Thorsten and Swaminathan, Adith and Schnabel, Tobias},
  title = {Unbiased Learning-to-Rank with Biased Feedback},
  year = {2017},
  isbn = {9781450346757},
  publisher = {Association for Computing Machinery},
  address = {New York, NY, USA},
  url = {https://doi.org/10.1145/3018661.3018699},
  doi = {10.1145/3018661.3018699},
  booktitle = {Proceedings of the Tenth ACM International Conference on Web Search and Data Mining},
  pages = {781–789},
  numpages = {9},
  location = {Cambridge, United Kingdom},
  series = {WSDM '17}
}

@article{gebru2021datasheets,
  author = {Gebru, Timnit and Morgenstern, Jamie and Vecchione, Briana and Vaughan, Jennifer Wortman and Wallach, Hanna and III, Hal Daum\'{e} and Crawford, Kate},
  title = {Datasheets for datasets},
  year = {2021},
  issue_date = {December 2021},
  publisher = {Association for Computing Machinery},
  address = {New York, NY, USA},
  volume = {64},
  number = {12},
  issn = {0001-0782},
  url = {https://doi.org/10.1145/3458723},
  doi = {10.1145/3458723},
  journal = {Commun. ACM},
  month = nov,
  pages = {86–92},
  numpages = {7}
}

@inproceedings{kirchenbauer2023watermark,
  title = {A Watermark for Large Language Models},
  author = {Kirchenbauer, John and Geiping, Jonas and Wen, Yuxin and Katz, Jonathan and Miers, Ian and Goldstein, Tom},
  booktitle = {Proceedings of the 40th International Conference on Machine Learning},
  pages = {17061--17084},
  year = {2023},
  editor = {Krause, Andreas and Brunskill, Emma and Cho, Kyunghyun and Engelhardt, Barbara and Sabato, Sivan and Scarlett, Jonathan},
  volume = {202},
  series = {Proceedings of Machine Learning Research},
  month = {23--29 Jul},
  publisher = {PMLR},
  url = {https://proceedings.mlr.press/v202/kirchenbauer23a.html},
}

@inproceedings{dai2024bias,
  author = {Dai, Sunhao and Xu, Chen and Xu, Shicheng and Pang, Liang and Dong, Zhenhua and Xu, Jun},
  title = {Bias and Unfairness in Information Retrieval Systems: New Challenges in the LLM Era},
  year = {2024},
  isbn = {9798400704901},
  publisher = {Association for Computing Machinery},
  address = {New York, NY, USA},
  url = {https://doi.org/10.1145/3637528.3671458},
  doi = {10.1145/3637528.3671458},
  booktitle = {Proceedings of the 30th ACM SIGKDD Conference on Knowledge Discovery and Data Mining},
  pages = {6437–6447},
  numpages = {11},
  location = {Barcelona, Spain},
  series = {KDD '24}
}

@inproceedings{zhou2025sourceecho,
  author = {Zhou, Yuqi and Dai, Sunhao and Pang, Liang and Wang, Gang and Dong, Zhenhua and Xu, Jun and Wen, Ji-Rong},
  title = {Exploring the Escalation of Source Bias in User, Data, and Recommender System Feedback Loop},
  year = {2025},
  isbn = {9798400715921},
  publisher = {Association for Computing Machinery},
  address = {New York, NY, USA},
  url = {https://doi.org/10.1145/3726302.3729972},
  doi = {10.1145/3726302.3729972},
  booktitle = {Proceedings of the 48th International ACM SIGIR Conference on Research and Development in Information Retrieval},
  pages = {1676–1686},
  numpages = {11},
  location = {Padua, Italy},
  series = {SIGIR '25}
}

@inproceedings{xu2024invisiblerelevance,
  author = {Xu, Shicheng and Hou, Danyang and Pang, Liang and Deng, Jingcheng and Xu, Jun and Shen, Huawei and Cheng, Xueqi},
  title = {Invisible Relevance Bias: Text-Image Retrieval Models Prefer AI-Generated Images},
  year = {2024},
  isbn = {9798400704314},
  publisher = {Association for Computing Machinery},
  address = {New York, NY, USA},
  url = {https://doi.org/10.1145/3626772.3657750},
  doi = {10.1145/3626772.3657750},
  booktitle = {Proceedings of the 47th International ACM SIGIR Conference on Research and Development in Information Retrieval},
  pages = {208–217},
  numpages = {10},
  location = {Washington DC, USA},
  series = {SIGIR '24}
}

@techreport{gyongyi2005webspam,
  number = {2004-25},
  month = {March},
  author = {Zoltan Gyongyi and Hector Garcia-Molina},
  title = {Web Spam Taxonomy},
  type = {Technical Report},
  publisher = {Stanford},
  institution = {Stanford InfoLab},
  year = {2004},
  url = {http://ilpubs.stanford.edu:8090/646/},
}

@misc{sadasivan2023impossible,
  title = {Can {AI}-Generated Text be Reliably Detected?},
  author = {Vinu Sankar Sadasivan and Aounon Kumar and Sriram Balasubramanian and Wenxiao Wang and Soheil Feizi},
  year = {2024},
  url = {https://openreview.net/forum?id=NvSwR4IvLO}
}

@inproceedings{su2025corpus,
  title = "Corpus Poisoning via Approximate Greedy Gradient Descent",
  author = "Su, Jinyan and Nakov, Preslav and Cardie, Claire",
  editor = "Che, Wanxiang and Nabende, Joyce and Shutova, Ekaterina and Pilehvar, Mohammad Taher",
  booktitle = "Findings of the Association for Computational Linguistics: ACL 2025",
  month = jul,
  year = "2025",
  address = "Vienna, Austria",
  publisher = "Association for Computational Linguistics",
  url = "https://aclanthology.org/2025.findings-acl.222/",
  doi = "10.18653/v1/2025.findings-acl.222",
  pages = "4274--4294",
  ISBN = "979-8-89176-256-5",
}

@inproceedings{chen-etal-2024-spiral,
  title = "Spiral of Silence: How is Large Language Model Killing Information Retrieval?{---}{A} Case Study on Open Domain Question Answering",
  author = "Chen, Xiaoyang and He, Ben and Lin, Hongyu and Han, Xianpei and Wang, Tianshu and Cao, Boxi and Sun, Le and Sun, Yingfei",
  editor = "Ku, Lun-Wei and Martins, Andre and Srikumar, Vivek",
  booktitle = "Proceedings of the 62nd Annual Meeting of the Association for Computational Linguistics (Volume 1: Long Papers)",
  month = aug,
  year = "2024",
  address = "Bangkok, Thailand",
  publisher = "Association for Computational Linguistics",
  url = "https://aclanthology.org/2024.acl-long.798/",
  doi = "10.18653/v1/2024.acl-long.798",
  pages = "14930--14951",
}

\end{document}